\documentclass[twocolumn,english,prl,showpacs,aps]{revtex4-1}
\usepackage[T1]{fontenc}
\usepackage[latin9]{inputenc}
\usepackage{amssymb}
\usepackage{times}
\usepackage{graphicx}
\usepackage{amsmath}
\usepackage{pstricks}
\usepackage{color}
\usepackage[normalem]{ulem}
\usepackage{xcolor}

\newcommand\redsout{\bgroup\markoverwith{\textcolor{red}{\rule[0.5ex]{2pt}{0.4pt}}}\ULon}

\begin{document}
\title{Condensate losses and oscillations induced by Rydberg atoms}

\author{
Tomasz~Karpiuk,$^{1}$ 
Miros{\l}aw~Brewczyk,$^{1}$ 
and Kazimierz~Rz\k{a}\.{z}ewski$^{2}$
}
\affiliation{
\mbox{$^1$Wydzia{\l} Fizyki, Uniwersytet w Bia{\l}ymstoku, ul. Cio{\l}kowskiego 1L, 15-245 Bia{\l}ystok, Poland}
\mbox{$^2$Center for Theoretical Physics PAN, Al. Lotnik\'ow 32/46, 02-668 Warsaw, Poland}  
}

\author{
Anita~Gaj,$^{3}$
Alexander~T.~Krupp,$^{3}$  
Robert~L\"{o}w,$^{3}$ 
Sebastian~Hofferberth,$^{3}$ 
and Tilman~Pfau$^{3}$
}
\affiliation{
\mbox{$^3$5. Physikalisches Institut and Center for Integrated Quantum Science and Technology IQST,}\\ 
\mbox{Universit\"at Stuttgart, Pfaffenwaldring 57, 70569 Stuttgart, Germany}
}

\begin{abstract}

We numerically analyze the impact of a single Rydberg electron onto a Bose-Einstein condensate. Both $S-$ and $D-$ Rydberg states are studied. The radial size of $S-$ and $D-$states are comparable, hence the only difference is due to the angular dependence of the wavefunctions. We find the atom losses in the condensate after the excitation of a sequence of Rydberg atoms. Additionally, we investigate the mechanical effect in which the Rydberg atoms force the condensate to oscillate. Our numerical analysis is based on the classical fields approximation. Finally, we compare numerical results to experimental data.

\end{abstract}


\maketitle

\section{Introduction}

Effects related to the interaction of a highly excited electron in a Rydberg atom with neutral atoms have been extensively studied for many decades \cite{Lebedev1995}. First seminal experiments were performed by Amaldi and Segr\`{e} \cite{Aml34,AS34I}, who studied Rydberg absorption lines for principal quantum number $n \leq 30$ of sodium atoms immersed in hydrogen background gas, with the hydrogen atoms playing the role of perturbers for the Rydberg electron. The important finding of these experiments was the shift of the absorption lines proportional to the density of the background gas. The high pressure of the perturbing gas (up to one
atmosphere) resulted in a large number of perturbing atoms (of the order of $10^4$) being in the volume of the Rydberg electron wavefunction. Surprisingly, the shift of the absorption lines can
be both negative or positive, depending on the kind of perturbing gas. The explanation of this observation was given by Fermi who described the interaction between the Rydberg electron and the atom introducing the pseudopotential \cite{Frm34,Frm36}.

This concept of applying the pseudopotential to the Rydberg electron-atom interaction \cite{Omont1977} led to the
prediction of ultralong-range Rydberg molecules \cite{GDS00}, which consist of a Rydberg atom and at least one ground state atom bound to the Rydberg electron by low-energy scattering. Rb$_2$ Rydberg molecules were experimentally realized in \cite{BBN09}. 
Since then, diatomic homonuclear Rydberg molecules were realized using
various atoms and Rydberg states like Rb in D-states \cite{KGB14,AMR14} and P-states \cite{BCB13};
Cs in S-states \cite{TRB12} and P-states \cite{SMD15}; Sr in S-states \cite{DAD15}. Recently, ultralong-range
Rydberg molecules bound by mixed singlet and triplet scattering have been predicted
\cite{AMR14b} and experimentally realized in \cite{SMD15,BGW16,NTE16b}. Coherent creation and breaking of molecular bonds has been shown in  \cite{BNB10}.
The
scattering resonance present for e-Rb p-wave scattering \cite{HGS02} led to the observation
of butterfly state Rydberg molecules \cite{NTE16} and Rydberg molecules bound by quantum
reflection \cite{BBN10}. Trilobite molecules possessing large permanent electric dipole moment has been observed in \cite{LPR11,Booth99}. 
Polyatomic Rydberg molecules with up to four ground state perturbers located inside the Rydberg electron orbit has been observed in \cite{GKB14}.

A Rydberg electron can interact with even thousands of ground state atoms
at high densities.
An experiment studying the coupling of the Rydberg electron to a Bose-Einstein condensate was performed \cite{LossesNature}. Large principal quantum numbers of the Rydberg states were considered allowing these states to have the size of a few micrometers, comparable to the size of the condensate. Hence, the Rydberg electron was interacting with a large (as in \cite{Aml34}) number of condensate atoms being within the extent of its orbit. As opposed to the conditions of the experiment of Ref. \cite{Aml34}, the Rydberg electron now interacts with the condensed atoms what makes a huge difference. The coupling of a single Rydberg electron to the condensate is surprisingly strong in comparison to the analogous coupling of the ionic impurity. The electron can excite the condensate atoms leading to the significant depletion of the condensate. The electron can as well induce the collective oscillations of the whole condensate \cite{LossesNature}.
Recently, Rydberg spectroscopy of a single Rydberg atom excited
in a BEC revealed the impact of the p-wave shape resonance onto the spectral
profile \cite{SLN16}. Furthermore, ultracold chemical reactions of a single Rydberg atom
excited in a BEC were studied in \cite{SLE16}.

In Ref. \cite{Karpiuk15} we have developed a stochastic model which describes the process of a creation of Rydberg atoms in the Bose-Einstein condensate and the following interaction of Rydberg electron with the condensed atoms. Our simulations \cite{Karpiuk15} show the agreement with the experimental results in the part regarding the condensate losses due to thermal depletion in the case of $S-$Rydberg states \cite{LossesNature}. Here, we extend our model to the $D-$Rydberg states and study also the condensate oscillations caused by the creation of a sequence of Rydberg atoms in the condensate.

\section{Theoretical model}

The single electron of a Rydberg atom polarizes near-by atoms. 
The atoms become electric dipoles. 
The interaction between an electric dipole and a charge is of short range type 
$\sim 1/r^4$.
To describe the interaction 
between the electron and the surrounding ground state atoms we use 
a pseudopotential \cite{Frm34,Frm36}.
In the simplest case only s-wave partial wave is considered and this potential takes the following form
\begin{equation}
 V_{Ryd}(\vec{r}^{\,\prime}) = \frac{2\pi \hbar^2 a(k^{\prime})}{m_e} |\Psi_{Ryd}(\vec{r}^{\,\prime})|^2 \,,
\end{equation}
where $\Psi_{Ryd}(\vec{r}^{\,\prime})$ is the Rydberg electron wavefunction, $m_e$ is the electron mass,
$\vec{r}^{\,\prime}$ is the position of electron with respect to the center of Rydberg atom and
$a(k^{\prime})$ denotes the electron-atom $s$-wave scattering length.
\begin{equation}
 a(k^{\prime}) = a_e + \frac{\hbar^2 \alpha}{e^2 a_0^2 m_e} k^{\prime} \,,
 \label{akprime}
\end{equation}
where $a_0$ is Bohr radius, $a_e$ is the zero-energy scattering length, 
$\alpha$ is the polarizability, $e$ is the elementary charge and 
$k^{\prime}$ is the electron wave number. 

In a more sophisticated approach also p-wave partial wave is taken into account \cite{Omont}.
Now, the Rydberg potential takes the following form
\begin{multline}
 V_{Ryd}(\vec{r}^{\,\prime}) = 2\pi A_0^{S,T}(k^{\prime}) |\Psi_{Ryd}(\vec{r}^{\,\prime})|^2 \\
 + 6\pi A_1^{S,T}(k^{\prime}) |\nabla \Psi_{Ryd}(\vec{r}^{\,\prime})|^2\,,
\end{multline}
where $A_l^{S,T}(k^{\prime}) = -\tan{\delta_l^{S,T}(k^{\prime})}/{k^{\prime}}^{2l+1}$
and $\delta_l^{S,T}(k^{\prime})$ are the scattering phase shifts. $S$ and $T$
refer to the singlet and triplet channels, respectively. Here we are interested in the triplet case.
The values of the scattering phase shifts are taken from \cite{phaseShifts}.

There is also the second charge coming from the nucleus.
We neglect this ingredient because a reduced mass of the neutral atom and the positive ion 
is about five orders of magnitude larger. This way the interaction energy is much smaller
than in case of the electron.

To describe the effect of the Rydberg electron on the condensate, the 
pseudopotential is introduced as an additional term in the Gross-Pitaevskii 
equation
\begin{eqnarray}
 i\hbar \frac{\partial}{\partial t} \Psi(\vec{r},t) &=& \left[
-\frac{\hbar^2}{2m}\nabla^2
+ V_{trap}(\vec{r})
+ g|\Psi(\vec{r},t)|^2
\right. \nonumber  \\
&&\left.
+ f(t)\;V_{Ryd}(\vec{r}-\vec{R})
\right] \Psi(\vec{r},t)  \,,
\end{eqnarray}
where $\Psi(\vec{r},t)$ is the condensate wave function, 
$g$ is the coupling constant for neutral atoms and 
$f(t)$ is a function which takes the values $1$ or $0$, depending on 
whether the Rydberg atom is present or not.
 It is important to appropriately model the excitation process 
in order to reproduce the experimental findings.
The excitation process depends both on time and position.

To find the position where the Rydberg atom is excited we use the following procedure.
We choose a position of a possible excitation according to the density 
distribution of neutral atoms $\rho(\vec{r},t)=|\Psi(\vec{r},t)|^2$. 
Then we draw a random number between $0$ and $1$ and compare to the excitation
probability $p(\vec{R},t)$ to determine if an excitation indeed takes place.
The probability to find an atom at position $\vec{R}$ in the Rydberg state 
is given by
\begin{equation}
 p(\vec{R},t) = 
\frac{\Omega_R^2}{\Omega^2(\vec{R},t)}\sin^2{\left[\Omega(\vec{R},t) \, t /2 
\right]}\,,
\label{Prob}
\end{equation}
where $\Omega_{R}$ is the single atom Rabi frequency in the vacuum. In the presence of
neutral atoms the Rydberg level gets shifted and a local Rabi frequency is given by
\begin{equation}
 \Omega(\vec{R},t) = \sqrt{\Omega_R^2 + \Delta^2(\vec{R},t)}\,,
\label{Omega}
\end{equation}
where $\Delta(\vec{R},t)$ is a local detuning. This quantity reads
\begin{equation}
 \Delta(\vec{R},t) = \Delta \omega_{L} - \delta E (\vec{R},t) / \hbar \,,
\end{equation}
where $\Delta \omega_{L}$ is the laser detuning from the vacuum Rydberg level,
$\delta E (\vec{R},t)$ is the shift of the Rydberg level caused by the neutral 
atoms. This shift, called the pressure effect \cite{Frm34,Aml34}, takes the following form
\begin{equation}
 \delta E (\vec{R},t) = \int V_{Ryd}(\vec{r} - \vec{R}\,)\;  \rho(\vec{r},t)\; 
d^3 r \,.
\label{PrssEff}
\end{equation}
As it was already mentioned we draw a random number between $0$ and $1$ and compare it
to the excitation probability $p(\vec{R},t)$.
If this random number is smaller than $p(\vec{R},t)$, $f(t)$ changes its value from $0$ to $1$.
If not,  another position is picked randomly and the procedure 
is repeated until either an excitation occurs or the number of trials reaches $N$, the number of atoms in the sample.
If the latter happens no Rydberg excitation occurs during this time step and we 
advance by a time-step of $\Delta t = 1/\Gamma$. 
Once created, the interaction of the Rydberg electron with the condensate 
is limited by a mean lifetime of Rydberg states $\tau$.

The rate $\Gamma$ we try to excite the Rydberg atom is also very important to
obtain results comparable with the experiment.
The coupling of the Rydberg electron to the condensate causes 
a process which identifies (i.e. measures) the position of the Rydberg atom.
At this point, the coherent evolution of the excitation, 
described by $p(\vec{R},t)$, stops.
We assume that this process originates from the elastic scattering of 
the Rydberg electron at the atoms in the condensate.
The scattering rate
\begin{equation}
\Gamma_{scat} = \int I(\vec{r}-\vec{R}) \sigma(\Omega,k^{\prime}) \rho(\vec{r},t) \; d^3 r \; d \Omega
\label{gammaInt}
\end{equation}
where $I(\vec{r^{\,\prime}}) = v(r^{\prime}) |\Psi_{Ryd}(\vec{r}^{\,\prime})|^2$ is the flux density, 
$v(r^{\prime}) = \sqrt{2 E_{kin}(r^{\prime}) / m_e}$ is 
the electron velocity and $\sigma(\Omega,k^{\prime})=a^2(k^{\prime})$ is the differential scattering cross section.
The semiclassical approximation for the kinetic energy of the Rydberg electron gives
\begin{equation}
E_{kin}(r^{\prime}) = 
- \frac{e^2}{2 a_0} \frac{1}{n^2}
+ \frac{e^2}{r^{\prime}}\,.
\label{ekinr}
\end{equation}
On the other hand
\begin{equation}
 E_{kin} = \frac{\hbar^2 {k^{\prime}}^2}{2m}\,.
 \label{ekink}
\end{equation}
Combining equations (\ref{ekinr}) and (\ref{ekink}) we can express
$k^{\prime}$ in terms of $r^{\prime}$ as
\begin{equation}
 k^{\prime}(r^{\prime}) = \frac{2m}{\hbar^2} \sqrt{- \frac{e^2}{2 a_0} \frac{1}{n^2}
+ \frac{e^2}{r^{\prime}}}\,.
\end{equation}
This way the differential scattering cross section $\sigma(k^{\prime}(r^{\prime}))$ 
in the integral (\ref{gammaInt}) becomes $r^{\prime}$ dependent.
The value of the scattering rate, calculated using the formula (\ref{gammaInt}), depends on where the Rydberg atom is
created. It changes from $0$ while it is created at the border of the condensate to about $8$ MHz in the center.

The time-dependent potential of the Rydberg atom heats the condensate and some ground state atoms are promoted
to the thermal cloud. These are called losses and were measured in \cite{LossesNature}. 
To find losses in our model we use the classical field approximation. 
In the framework of this approach the condensate and the thermal cloud are identified by
the coarse graining the one-particle density matrix \cite{Goral,Brewczyk}. 
In the present work the one-particle density matrix is coarse grained by the column integration \cite{Karpiuk}.
\begin{equation}
\bar{\rho}(x,z,x',z';t) = \frac{1}{N}  \int dy \, \Psi(x,y,z,t) \, \Psi^*(x',y,z',t)  \,.
\label{average}
\end{equation}
After spectral decomposition \cite{Penrose}, the fraction of the condensed atoms are given by a dominant eigenvalue.
The amount of remaining atoms measures the losses.

\section{Condensate losses}

One of the main outcomes of article~\cite{LossesNature} are resonance lines
appearing in the relative BEC atom number after a series of Rydberg atoms 
was excited. These lines were measured for the Rydberg atoms in three different
states, namely for $(n=110,\,l=0,\,m_l=0)$, $(n=106,\,l=2,\,m_l=0)$ and 
$(n=106,\,l=2,\,m_l=2)$. 
In the further discussion we will call them $110S$, $106D0$ and $106D2$.
The experimental sequence was as follows.
First, a BEC of $^{87}$Rb atoms was created.
The number of atoms in the condensate was about $1\times 10^5$ plus some amount of atoms in the thermal cloud.
Then a series of Rydberg atoms was excited by two photon excitation 
process. 
To avoid an unnecessary shaking of condensate the red laser was switched on
by ramping adiabatically from zero to its maximum value in $24$ ms.
Then the power of the red laser was kept at its maximum for additional $14$ ms.
Finally, the condensate was released from the trap and time of flight (TOF) imaging was performed after $50\,$ms.
From these images both the condensate losses and the condensate oscillations were extracted. 

\begin{figure}[thb]
\includegraphics[width=7.7cm]{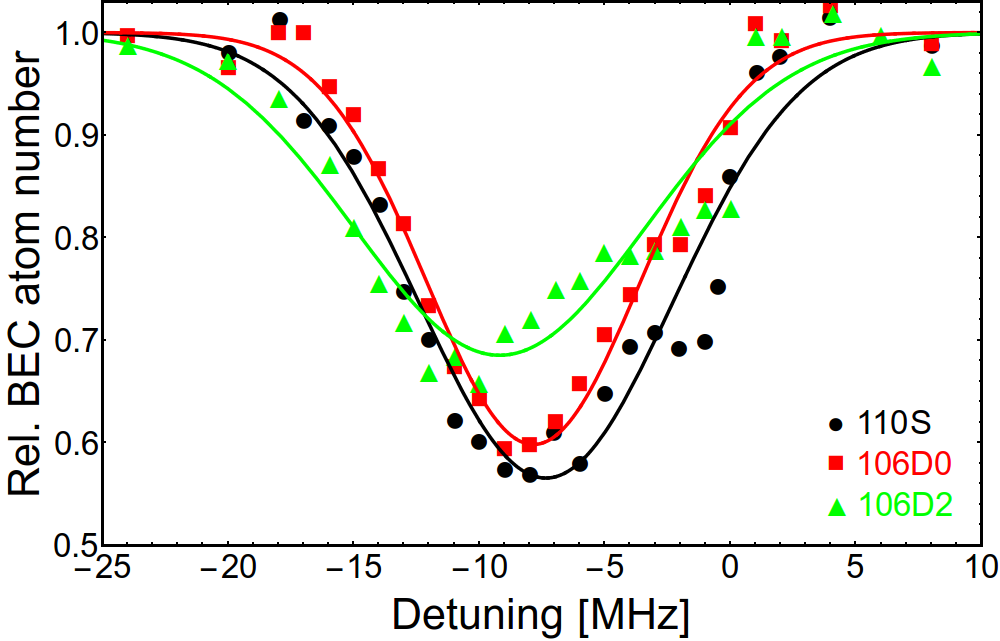}
\caption{(color online). 
The experimental relative BEC atom number versus the 
detuning from the non-interacting Rydberg level. 
The red color, the green color and the blue color correspond to 
the $110S$ state, the $106D0$ state and the $106D2$ state respectively. 
The points are experimental results whereas the lines are the Gaussian
fits.
}
\label{Exp}
\end{figure}

Experimental results of losses are shown in Fig.~\ref{Exp}  
 as black disks ($110S$), red squares ($106D0$), and green triangles ($106D2$).
 In all cases the Rabi frequency $\Omega_R=408$Hz. 
Each point is an average result of twenty
 repetitions of the experiment.

In case of $110S$ and $106D0$ the experimental lines are quite symmetric. In 
the last case the line is clearly asymmetric. Despite this fact we do the Gaussian
fits and we extract the minimum value, the full width at half magnitude (FWHM) and 
the position of each line. The results are collected in Table \ref{fitsExp}.

\begin{table}[thb]
 \begin{tabular}{|c||c|c|c|}
 \hline
 State  & $n_0$             & FWHM [MHz]     & $\Gamma_0$ [MHz]     \\
 \hline\hline
 110S   & $0.565 \pm 0.018$ & $11.9 \pm 0.6$ & $-7.3 \pm 0.2$ \\
 \hline
 106D0  & $0.598 \pm 0.012$ & $10.0 \pm 0.4$ & $-7.83 \pm 0.15$ \\
 \hline
 106D2  & $0.685 \pm 0.015$ & $13.6 \pm 0.9$ & $-9.2 \pm 0.3$ \\
 \hline
 \end{tabular}
 \caption{The values of parameters of the Gaussian fits corresponding to the 
 experimental data.
 }
 \label{fitsExp}
\end{table}

Before we will try to compare the theoretical results with the experiment
we would like to discuss how the parameters of our model influence the shape 
of the resonance lines. There are three parameters we are going to alter, namely:
the scattering rate $\Gamma$, the Rabi frequency $\Omega_R$ and the lifetime
$\tau$.

Simulations show that the condensate fraction goes down when the scattering rate grows.
This result is easy to understand. Increasing the scattering rate we increase 
the frequency we try to excite the Rydberg atom. This way more
Rydberg atoms get excited. This in a
straight way leads to increase of losses, i.e., the reduction of the condensate fraction.
While the losses are affected, the line width and the line 
position remain almost unchanged. 

Similar analysis is done for different values of the Rabi frequency. 
The increase of the Rabi 
frequency leads to the increase of losses. This is caused by increase
of the probability to excite the Rydberg atom. This fact is obvious from the 
formula (\ref{Prob}) written for small times which is the case in our simulations.
The probability is then proportional to $\Omega^2_R\,t^2/4$.
Two other parameters, the FWHM and the line
position, remain almost unaffected as in the previous case.

The lifetime of the Rydberg state influences the resonance lines in the strongest way. 
Together with growth of the lifetime both losses and FWHM are growing. 
The increase in losses is very strong. 
Surprisingly this increase in losses is accompanied by the reduction of the number of
Rydberg atoms created during the excitation sequence. 
So, longer the life time, the smaller
number of Rydberg atoms is created during the same time. 
Longer the lifetime, longer the total time when Rydberg atoms interact with the neutral atoms and shorter the total
time when there is no Rydberg atom in the system.
The lines positions are almost unaffected.

We try to recover the experimental resonance lines using our procedure. 
The initial number of atoms in the unperturbed condensate is $1.2\times 10^5$ in our simulations.
Then we prepare a cloud of atoms at the final temperature containing $1\times 10^5$ of atoms
in the condensate and $0.2\times 10^5$ of atoms in the thermal cloud and follow the experimental
procedure described earlier in this section.
If we use directly the experimental values of parameters only the $110S$ state line agrees with the 
experiment in terms of the maximum losses. 
In case of $106D0$ state and $106D2$ state losses are to high. 
So we tune the scattering rate and the lifetime to get the maximum losses close to the experimental one. 
We always use the Rabi frequency $\Omega_R = 408$ Hz. 
So if we chose the scattering rate $\Gamma = 5$ MHz,
the lifetime $\tau=4.8$ $\mu$s for the $110S$ state, $\Gamma = 1.25$ MHz, $\tau=3.2$ $\mu$s for the $106D0$ state
and $\Gamma = 1.25$ MHz, $\tau=3.2$ $\mu$s for the $106D2$ state
we are close to the experimental maximum losses. 
The chosen very long experimental sequence minimizes the influence of the red laser 
light onto the many-body dynamics of the BEC with respect to the work presented in \cite{LossesNature,Karpiuk15}. 
However, at the same time, there are complex chemical processes which could take place on such long time scale \cite{SLE16,NTM15b}. 
Given the limitations in our knowledge about the exact processes, we use a simplified description
introducing an effective lifetime as a first approach, which results in a  qualitative agreement
with the experimental data.

\begin{figure}[thb]
\centerline{
\includegraphics[width=8.5cm]{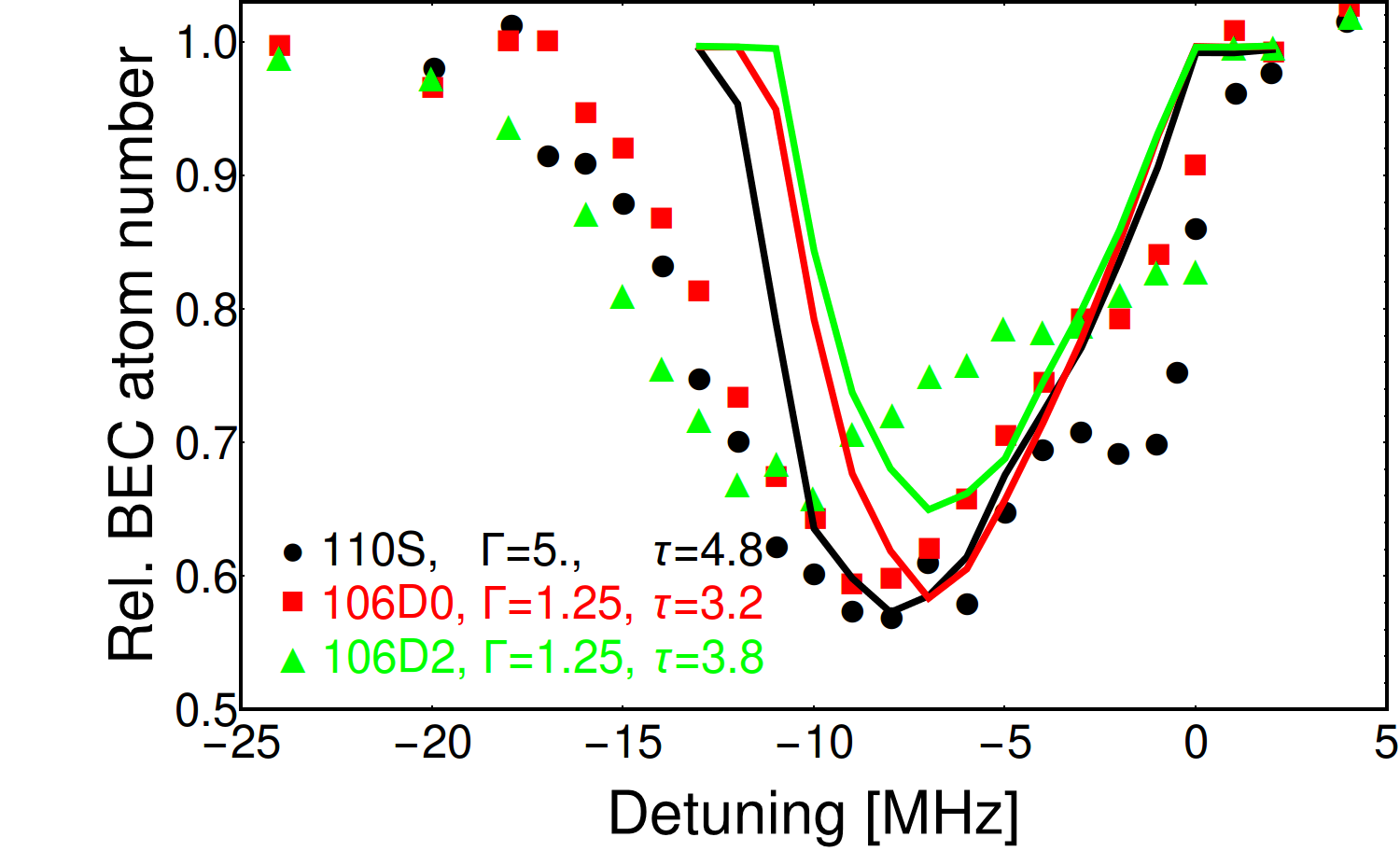}
\phantom{aaaaa}
}
\caption{(color online). 
The comparison between the experiment and the theory in terms of the relative BEC
atom number. Points show experimental values whereas lines theoretical one. 
The $110S$ state is shown in black, the $106D0$ state is shown as red and $106D2$ state is
shown as green respectively.
}
\label{N110S}
\end{figure}

The comparison between the experiment and the theory is shown in Fig.~\ref{N110S}.
The experimental results are shown as the black disks, the red squares and the green triangles for
$110S$, $106D0$ and $106D2$ state, respectively. The theoretical results are drawn as
lines.
The condensate fraction at maximum losses, the line width and the line position are extracted
by Gaussian fits and collected in Table \ref{fitsTry}.

We are able to change the maximum losses a lot by changing the Rabi frequency, 
the scattering rate or the life time.
This way we can get lines with the magnitude and position close to experimental
results.
Unfortunately there is no way to change the width
of the lines significantly using these three parameters.
The theoretical width is smaller than the experimental result.
We have checked that the width of the line can be increased by increasing the number of atoms. 

\begin{table}[thb]
 \begin{tabular}{|c||c|c|c|}
 \hline
 State  & $n_0$             & FWHM [MHz]     & $\Gamma_0$ [MHz]     \\
 \hline\hline
 110S   & $0.56 \pm 0.02$ & $7.4 \pm 0.5$ & $-7.1 \pm 0.2$ \\
 \hline
 106D0  & $0.575 \pm 0.018$ & $6.5 \pm 0.3$ & $-6.44 \pm 0.13$ \\
 \hline
 106D2  & $0.635 \pm 0.017$ & $6.4 \pm 0.4$ & $-6.22 \pm 0.15$ \\
 \hline
 \end{tabular}
 \caption{The values of parameters of the Gaussian fits for three considered states.}
 \label{fitsTry}
\end{table}

\section{Condensate oscillations}

\begin{figure}[thb]
\includegraphics[width=8.cm]{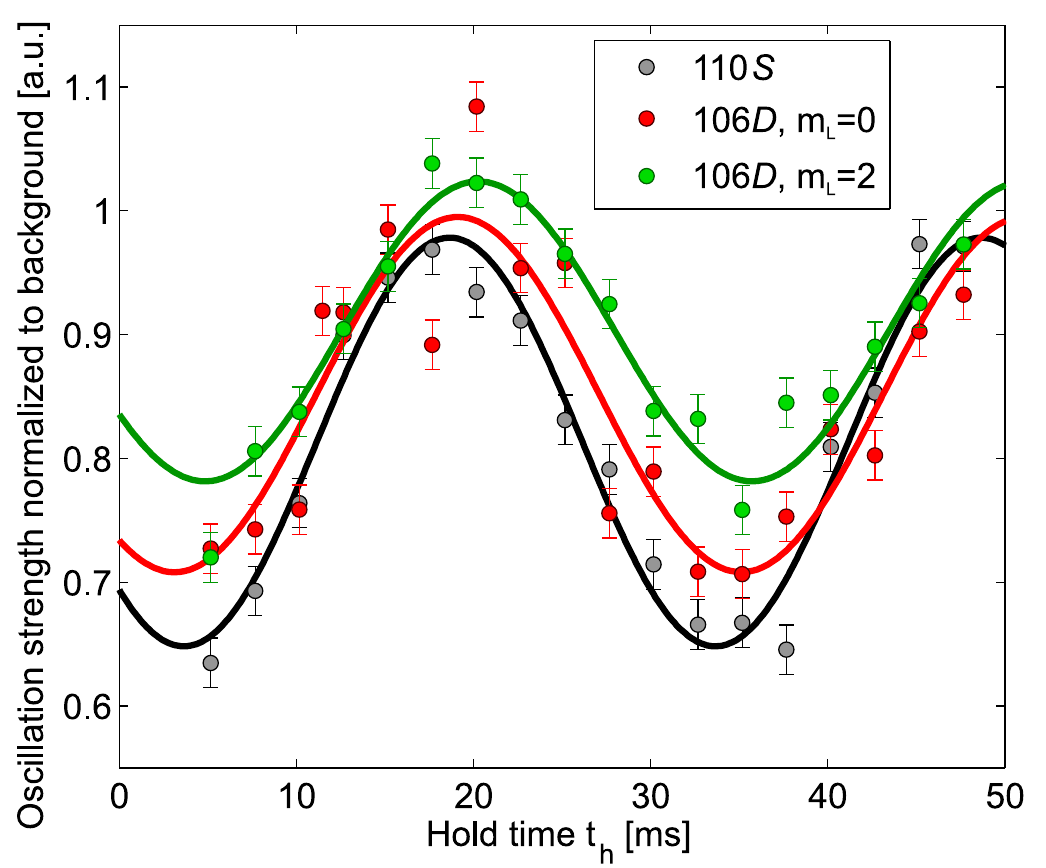}
\caption{(color online). 
The experimental oscillation strength of the condensate versus the hold time after the excitation sequence 
for $\Gamma=-8.5$MHz. Black, red and green points correspond to $110S$, $106D0$ and $106D2$ state respectively. 
}
\label{OscFigExp}
\end{figure}

After the series of Rydberg atoms is excited the remaining condensate undergoes
oscillations.
So there is a mechanical effect visible. 
This effect depends on the state of Rydberg atom which is used.
To describe this oscillations quantitatively we measure the width of the condensate
in the radial direction. 
This measurement is done for the detuning equal $-8.5$MHz which is on average close
to the detuning leading to the maximum losses for each state.
The measurement is done by fitting the following function to the radial and axial cuts of the measured column densities
\begin{equation}
 d(\alpha) = p_{1} \left(1-\frac{(\alpha-p_{2})^2}{{p_{3}^2}}\right)^{3/2}\,,
\end{equation}
where $\alpha$ is the radial or axial coordinate respectively, $p_1$ is the peak column density,
$p_2$ is the average position of the condensate and $p_3$ is the condensate width. 
From these fits we get $p_{3}$ for the radial direction. Then we calculate the radial width normalized 
to the unperturbed radial width and repeat this procedure for growing the hold time $t_h$.
This quantity versus the hold time $t_h$ is shown in Fig. \ref{OscFigExp}.

Then to extract the amplitude of oscillations and the frequency 
we fit the sinusoidal function to this data
\begin{equation}
 A \sin{(\omega\, t + \phi)} + B \,,
\end{equation}
where $A$ is the amplitude of oscillations, $\omega$ is the frequency of oscillations,
$\phi$ is the phase shift and $B$ is the average strength of oscillations.
The values of the quantities we are interested in are shown in Table \ref{OscParExp}. 

\begin{table}[thb]
 \begin{tabular}{|c||c|c|}
 \hline
 State  & $A/B\,[$\%$]$  & $\omega[2\pi$Hz$]$     \\
 \hline\hline
 110S   & $16.5 \pm 1.1$ & $33.3 \pm 0.8$ \\
 \hline
 106D0  & $14.3 \pm 2.1$ & $31.2 \pm 1.6$ \\
 \hline
 106D2  & $12.1 \pm 1.1$ & $32.4 \pm 1.1$ \\
 \hline
 \end{tabular}
 \caption{The relative amplitude and the frequency of oscillations for different
 states measured in the experiment (look (fig.~\ref{OscFigExp})).}
 \label{OscParExp}
\end{table}

The oscillation frequencies are close to the slow quadrupole oscillation frequency 
$\omega \approx \sqrt{5/2} \omega_z = 2\pi\,34.8$Hz. The oscillation amplitude is
strongest for the $110S$ and smallest for the $106D2$ state with the $106D0$ state
in between.

\begin{figure}[thb]
\centerline{
\includegraphics[width=8.5cm]{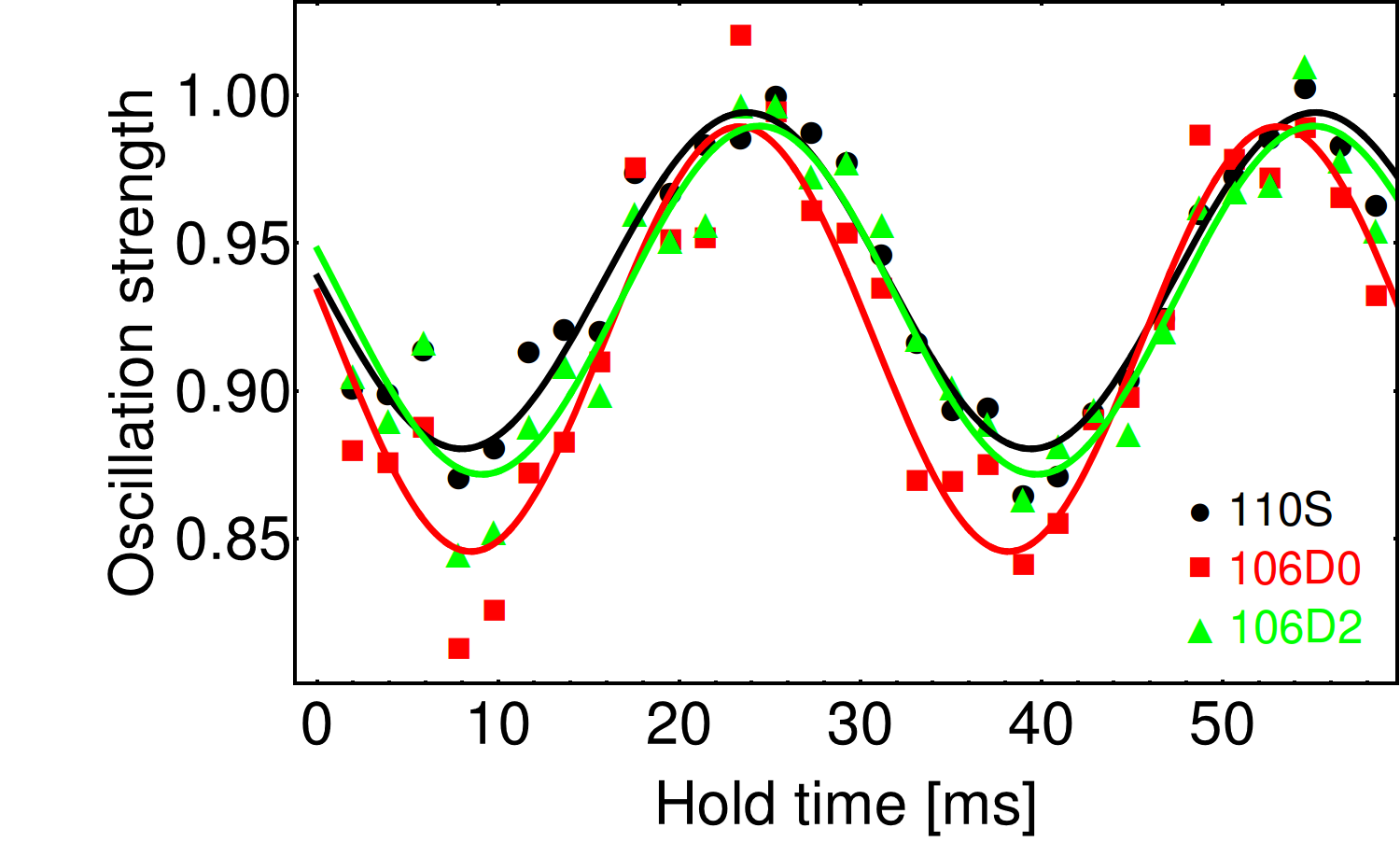}
\phantom{aaaaa}
}
\caption{(color online). 
The theoretical oscillation strength after the excitation sequence for $\Gamma=-8.0$MHz. 
The black disks, the red squares and the green triangles correspond to the $110S$ state,  
the $106D0$ state and the $106D2$ state respectively. 
}
\label{OscFigTry}
\end{figure}

We repeat the experimental procedure using our theoretical description.
Then we measure the condensate width in axial direction. 
This step is different than in the experiment. 
The width measurement in the experiment was done after the TOF expansion and
the radial width was bigger than the axial one and therefore more suitable to
perform measurements.
In the theoretical description we measure the width of the condensate inside 
the harmonic trap. The axial width is larger and this way better to do the width measurement.
In this case the oscillations are measured for the detuning equals $-8.0$MHz which
roughly corresponds to the maximum losses in the theoretical description.
The strength of oscillations is shown in Fig. \ref{OscFigTry}.

The parameters of the sinusoidal fit are collected in Table \ref{OscParTry}.

\begin{table}[thb]
 \begin{tabular}{|c||c|c|c|c|}
 \hline
 State  & $A/B\,[$\%$]$ & $\omega\,[2\pi$Hz$]$  \\
 \hline\hline
 110S   & $6.1 \pm 0.3$ & $31.8 \pm 0.6$ \\
 \hline
 106D0  & $7.8 \pm 0.6$ & $33.7 \pm 0.7$ \\
 \hline
 106D2  & $6.3 \pm 0.5$ & $32.7 \pm 0.7$ \\
 \hline
 \end{tabular}
  \caption{The relative amplitude and the frequency of oscillations for different
 states coming from the theory (look (fig.~\ref{OscFigTry})).}
 \label{OscParTry}
\end{table}

The main observation is that our theory can capture the mechanical effect.
The frequency of oscillations agrees very well with the experiment.
The phase of oscillations is very close to the experimental value.
Unfortunately the oscillation amplitude is roughly two times smaller than measured.
There are at least two reasons why this discrepancy may appear.
Firstly, in the experiment the amplitude is calculated with respect to the reference 
amplitude which comes from runs where the red laser is detuned from resonance.
The presence of the red laser introduces some small oscillations and affects in some
way the final result.
In the theoretical description we always divide the calculated amplitude by the unperturbed amplitude.
Secondly we measure the axial width inside the trap instead of the radial width after TOF.
Both effects can affect the final result and hence it is rather expected that we do not
see full quantitative agreement.
It has been shown in \cite{LossesNature} that the mechanical effect of the Rydberg atom on 
the BEC strongly depends on the principal quantum number $n$ of the excited Rydberg state.
Here, we show that many-body dynamics is essentially the same for two angular momenta states S and D,
which confirms the general applicability of the wavefunction imaging method in a BEC \cite{Karpiuk15}.

\section{Conclusions}

We have analyzed numerically the impact of the excitation of a sequence of Rydberg atoms on the Bose-Einstein condensate. 
We found that, by creating Rydberg atoms, the condensate is heated. 
Hence, the number of condensed atoms decreases. 
After the cloud of atoms is released from the trap, the thermal component is gone and the condensate losses can be measured. 
We calculated the final (i.e., when the excitation of Rydberg atoms is stopped) condensate fraction both for $S-$ and $D-$ Rydberg states. 
In fact, within the stochastic model we have developed we are able to obtain the full resonance curve. 
We find that the position of the center of the resonance line is determined solely by the total number of atoms in the sample 
whereas its width and depth strongly depend on the vacuum Rabi frequency, 
the scattering rate, and the lifetime of the Rydberg atom, i.e., on the parameters of our stochastic model. 
The scattering rate is calculated within the simplified semiclassical approach and the lifetime of the Rydberg atom is known from the measurement. 
We compared our numerical results to experimental data. 
There is a good agreement in the case of $S-$states. 
The good agreement for $D-$states is achieved  by use of an effective
 lifetime of the Rydberg atom. 
We also investigated the mechanical effect of Rydberg atoms on the condensate. 
We found the oscillations of the size of the condensate. 
The frequencies of these oscillations agree very well with experimentally measured while the amplitudes remain in qualitative agreement.

\section{Acknowledgments}
We are grateful to Mariusz Gajda, Tomasz Sowi\'{n}ski and Krzysztof Paw\l{}owski for helpful discussions. 
The work was supported by the (Polish) National Science Center Grant No. DEC-2012/04/A/ST2/00090 (T.K., M.B., and K.R.).
This work was supported by the Deutsche Forschungsgemeinschaft
(DFG) within the SFB/TRR21 and the
project PF 381/13-1. Parts of this work was also funded
by ERC under contract number 267100. S.H. acknowledges
support from DFG through the project HO 4787/1-1. We also acknowledge the European
Union H2020 FET Proactive project RySQ (grant N. 640378).

\bibliography{biblio}

%
%
%
%
%
%
%
%
%
%

\end{document}